\documentclass[prb,twocolumn,showpacs,preprintnumbers,amsmath,aps]{revtex4}
\usepackage{graphicx,hyperref}% Include figure files
\usepackage{bm}% bold math
\usepackage{subfigure}% bold math

\newcommand\tr{\text{Tr}}

\def\nbC{{\mathchoice {\setbox0=\hbox{$\displaystyle\rm C$}%
\hbox{\hbox to0pt{\kern0.4\wd0\vrule height0.9\ht0\hss}\box0}}
{\setbox0=\hbox{$\textstyle\rm C$}\hbox{\hbox
to0pt{\kern0.4\wd0\vrule height0.9\ht0\hss}\box0}}
{\setbox0=\hbox{$\scriptstyle\rm C$}\hbox{\hbox
to0pt{\kern0.4\wd0\vrule height0.9\ht0\hss}\box0}}
{\setbox0=\hbox{$\scriptscriptstyle\rm C$}\hbox{\hbox
to0pt{\kern0.4\wd0\vrule height0.9\ht0\hss}\box0}}}}
\def\nbQ{{\mathchoice {\setbox0=\hbox{$\displaystyle\rm
Q$}\hbox{\raise
0.15\ht0\hbox to0pt{\kern0.4\wd0\vrule height0.8\ht0\hss}\box0}}
{\setbox0=\hbox{$\textstyle\rm Q$}\hbox{\raise
0.15\ht0\hbox to0pt{\kern0.4\wd0\vrule height0.8\ht0\hss}\box0}}
{\setbox0=\hbox{$\scriptstyle\rm Q$}\hbox{\raise
0.15\ht0\hbox to0pt{\kern0.4\wd0\vrule height0.7\ht0\hss}\box0}}
{\setbox0=\hbox{$\scriptscriptstyle\rm Q$}\hbox{\raise
0.15\ht0\hbox to0pt{\kern0.4\wd0\vrule height0.7\ht0\hss}\box0}}}}
\def\nbT{{\mathchoice {\setbox0=\hbox{$\displaystyle\rm
T$}\hbox{\hbox to0pt{\kern0.3\wd0\vrule height0.9\ht0\hss}\box0}}
{\setbox0=\hbox{$\textstyle\rm T$}\hbox{\hbox
to0pt{\kern0.3\wd0\vrule height0.9\ht0\hss}\box0}}
{\setbox0=\hbox{$\scriptstyle\rm T$}\hbox{\hbox
to0pt{\kern0.3\wd0\vrule height0.9\ht0\hss}\box0}}
{\setbox0=\hbox{$\scriptscriptstyle\rm T$}\hbox{\hbox
to0pt{\kern0.3\wd0\vrule height0.9\ht0\hss}\box0}}}}
\def\nbS{{\mathchoice
{\setbox0=\hbox{$\displaystyle     \rm S$}\hbox{\raise0.5\ht0%
\hbox to0pt{\kern0.35\wd0\vrule height0.45\ht0\hss}\hbox
to0pt{\kern0.55\wd0\vrule height0.5\ht0\hss}\box0}}
{\setbox0=\hbox{$\textstyle        \rm S$}\hbox{\raise0.5\ht0%
\hbox to0pt{\kern0.35\wd0\vrule height0.45\ht0\hss}\hbox
to0pt{\kern0.55\wd0\vrule height0.5\ht0\hss}\box0}}
{\setbox0=\hbox{$\scriptstyle      \rm S$}\hbox{\raise0.5\ht0%
\hboxto0pt{\kern0.35\wd0\vrule height0.45\ht0\hss}\raise0.05\ht0%
\hbox to0pt{\kern0.5\wd0\vrule height0.45\ht0\hss}\box0}}
{\setbox0=\hbox{$\scriptscriptstyle\rm S$}\hbox{\raise0.5\ht0%
\hboxto0pt{\kern0.4\wd0\vrule height0.45\ht0\hss}\raise0.05\ht0%
\hbox to0pt{\kern0.55\wd0\vrule height0.45\ht0\hss}\box0}}}}
\def\nbZ{{\mathchoice {\hbox{$\sf\textstyle Z\kern-0.4em Z$}}
{\hbox{$\sf\textstyle Z\kern-0.4em Z$}}
{\hbox{$\sf\scriptstyle Z\kern-0.3em Z$}}
{\hbox{$\sf\scriptscriptstyle Z\kern-0.2em Z$}}}}
\begin{document}

\title{Same universality class for the critical behavior in and out of equilibrium in a quenched random field}

\author{Ivan Balog} \email{balog@ifs.hr}
\affiliation{LPTMC, CNRS-UMR 7600, Universit\'e Pierre et Marie Curie,
bo\^ite 121, 4 Pl. Jussieu, 75252 Paris c\'edex 05, France}
\affiliation{Institute of Physics, P.O.Box 304, Bijeni\v{c}ka cesta 46, HR-10001 Zagreb, Croatia}

\author{Matthieu Tissier} \email{tissier@lptmc.jussieu.fr}
\affiliation{LPTMC, CNRS-UMR 7600, Universit\'e Pierre et Marie Curie,
bo\^ite 121, 4 Pl. Jussieu, 75252 Paris c\'edex 05, France}

\author{Gilles Tarjus} \email{tarjus@lptmc.jussieu.fr}
\affiliation{LPTMC, CNRS-UMR 7600, Universit\'e Pierre et Marie Curie,
bo\^ite 121, 4 Pl. Jussieu, 75252 Paris c\'edex 05, France}

\date{\today}

\begin{abstract}

The random-field Ising model (RFIM) is one of the simplest statistical-mechanical models that captures the anomalous 
irreversible collective response seen in a wide range of physical, biological, or socio-economic situations in the presence of interactions 
and intrinsic heterogeneity or disorder. When slowly driven at zero temperature it can display  
an out-of-equilibrium phase transition associated with critical scaling (``crackling noise''), while it undergoes at equilibrium, 
under either temperature or disorder-strength changes, a thermodynamic phase transition. 
We show that the out-of-equilibrium and equilibrium critical behaviors are in the same universality class: they are controlled, 
in the renormalization-group (RG) sense, by the same zero-temperature fixed point. We do so by combining a field-theoretical 
formalism that accounts for the multiple metastable states and the exact (functional) RG. As a spin-off, 
we also demonstrate that critical fluids in disordered porous media are in the same universality class as the RFIM, thereby 
unifying a broad spectrum of equilibrium and out-of-equilibrium phenomena.

\end{abstract}


\maketitle

In the presence of both interactions and intrinsic heterogeneity or quenched disorder, a wide spectrum of systems display, 
when slowly driven by an external solicitation, discrete collective events, bursts, shocks, jerks, or avalanches, that span a 
broad range of sizes. The signature appears as a ``crackling noise''\cite{sethna01} and it can be found in quite different situations,
\cite{sethna01,fisher98,dahmen-zion} from Barkhausen noise in disordered magnets\cite{bertotti} to a variety of social and economic 
phenomena\cite{bouchaud13} in passing by capillary condensation in mesoporous materials\cite{hallock96,detcheverry04} 
or hysteresis and noise in disordered electron nematics in high-$T_c$ superconductors.\cite{nematic1,nematic2} 

It has been shown that a simple system such as the random-field Ising model (RFIM) has already 
all the required ingredients to display crackling noise.\cite{sethna93,dahmen96,sethna05} In this case, the latter results from the presence of an 
out-of-equilibrium critical point in the hysteretic response of the system to an infinitely slowly changed external field at zero temperature. 
The critical point separates a phase characterized by 
finite-size avalanches and a continuous hysteresis curve from a phase 
with a macroscopic avalanche and a discontinuous hysteresis curve. It requires tuning two control parameters, 
the disorder strength and the external magnetic field.

On the other hand, the RFIM has been for decades a model for equilibrium phase behavior in the presence of quenched 
disorder.\cite{imry-ma75,nattermann98}  For dimensions greater than $2$, the RFIM shows a phase transition and a critical 
point at fixed disorder strength when changing temperature or at fixed temperature when changing disorder strength.\cite{nattermann98} 

The puzzle we address and solve in the present work is the following: despite the fact that one is at equilibrium and the other is not, one 
is at zero external field and the other is not, and that they take place at different values of the disorder strength, the two types of critical points 
are characterized by critical exponents and  scaling functions that have been found very close in numerical simulations, within 
numerical accuracy.\cite{maritan94,perez04,liu09} (A similar observation concerning the critical exponents 
can be made from experiments but the uncertainties are much bigger.)

The theoretical tool for a proper resolution of this puzzle is the renormalization group (RG). The critical behaviors in and out of equilibrium 
are the same, and are therefore in the same universality class, \textit{iff} they are controlled by the same RG ``fixed point''. (A first piece of 
information is that the fixed points associated with both types of criticality occur at zero temperature where sample-to-sample fluctuations 
dominates over thermal ones;\cite{villain84,fisher86} however, this is a necessary but not sufficient condition.\cite{ledoussal-wiese})
A first attempt through an RG formalism was proposed on the basis of perturbation theory.\cite{dahmen96} However, the latter is known 
to seriously fail in the RFIM\cite{nattermann98,bricmont} and cannot provide a useful method. The route we follow here is based 
on the exact RG and builds on our recent work on the equilibrium behavior of the RFIM.\cite{tarjus04,tissier11,tissier12b}

Our demonstration proceeds in several steps. The first one is to replace the \textit{a priori} complex problem of following a 
history-dependent evolution among configurations, which results from the dynamics of the slowly driven RFIM, by one that is 
more readily tackled by statistical mechanical methods. The limit of interest is the adiabatic, or quasi-static, one in which the driving 
rate is vanishingly slow so that the system reaches a stationary state before being evolved again.\cite{sethna93} The trick is that, due to 
the ferromagnetic nature of the interactions in the RFIM and the properties of the zero-temperature 
relaxational dynamics (and the associated ``no-passing rule''\cite{middleton92,sethna93}), the configurations visited along the  
hysteresis loop correspond to \textit{extremal} states:\cite{guagnelli93,mlr10} for a given value of the 
applied magnetic field (in the language of magnetic systems), they correspond to the stationary states that have the largest local 
magnetization at each point (for the ``descending'' branch obtained by decreasing the magnetic field from a fully positively magnetized 
configuration, see Fig.~1) or the smallest one (for the ``ascending branch'' obtained by increasing the field from a fully negatively magnetized 
configuration). When the distribution of the random fields is continuous, 
which we shall consider, these extremal states are unique for a given realization of the disorder (with exceptional 
degeneracies).\cite{guagnelli93} One can then formulate \textit{a statistical mechanical treatment of the extremal 
states with no reference to dynamics and history}.
\begin{figure}[ht]
\includegraphics[width=\linewidth]{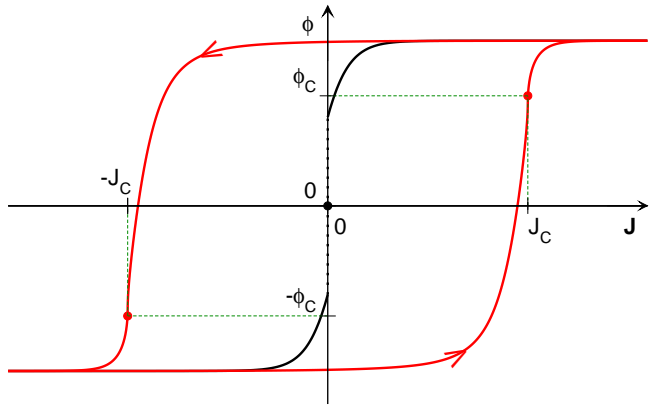}
\caption{
RFIM at zero temperature: Schematic illustration of the hysteresis loop (red) and of the equilibrium curve (blue) in the magnetization ($\phi$) 
versus applied magnetic field ($J$) representation. For the chosen value of the disorder strength, the system has two (symmetric) 
out-of-equilibrium critical points (red dots), but a first-order transition in equilibrium. }
\end{figure}

The model that we consider is the field-theoretical version of the RFIM with short-ranged interactions. The associated ``bare action'' 
(microscopic Hamiltonian) is
\begin{equation}
\begin{aligned}
\label{eq_ham_dis}
&S[\varphi;h+J]=  S_B[\varphi]-\int_{x} [h(x)+J(x)] \varphi(x) \,, \\&
S_B[\varphi]= \int_{x}\bigg\{\frac{1}{2}[\partial \varphi(x)]^2+ \frac{r}{2} \varphi(x)^2 + \frac{u}{4!} \varphi(x)^4 \bigg\},
\end{aligned}
\end{equation}
where $ \int_{x} \equiv \int d^d x$, $h(x)$ is a random ``source'' (a random magnetic field) and $J(x)$ is an external source (a magnetic field); 
the quenched random field is taken with a Gaussian distribution characterized by a zero mean and a variance 
$\overline{h(x)h(y)}= \Delta_B \delta^{(d)}( x-y)$.

At zero temperature, the stationary states under consideration are solutions of the stochastic field equation 
$\frac{\delta S[\varphi;h+J]}{\delta \varphi(x)}= 0$.
As discussed above, the states that are relevant for the quasi-static hysteresis curve are the two extremal solutions (Fig.~1).
 
The general recipe to build a generating functional from which one can derive all the needed correlation functions describing 
the extremal states is (i) to introduce a weighting factor with an auxiliary source linearly coupled to the $\varphi$ 
field to select the magnetization, and (ii) to consider copies or replicas of the original disordered system, each being 
independently coupled to distinct external sources.\cite{tissier11,tissier12b}  
The associated generating functional is then
\begin{equation}
\label{eq_generating_func1}
\begin{aligned}
\overline{\mathcal Z_h[\{ \hat{J}_a,J_a\}]}=&\int \prod_a \mathcal D\varphi_a \; \overline{\delta\left[\dfrac{\delta S_B[\varphi_a]}
{\delta \varphi_a}-h-J_a\right] }\;\times \\& \det\left[ \dfrac{\delta^2 S_B[\varphi_a]}{\delta \varphi_a \delta \varphi_a}\right] \exp  \int_{x} 
\hat{J}_a(x)  \varphi_a(x) \,,
\end{aligned}
\end{equation}
where the overline denotes an average over the Gaussian random field $h(x)$ and square brackets generically indicate functionals. 

Note that the above generating functional includes contributions from all solutions of the stochastic field equation (for each copy $a$). 
However, in the limit where all auxiliary sources $\hat{J}_a$ go to infinity, \textit{the dominant contribution is that of the extremal 
states}, with maximum magnetization when $\hat{J}_a \rightarrow +\infty$ and minimum one when $\hat{J}_a \rightarrow -\infty$. The 
correlation functions are then obtained by first differentiating $\log(\overline{\mathcal Z_h[\{ \hat{J}_a,J_a\}]})$ 
with respect to the $\hat{J}_a$'s 
and then taking the latter to infinity while considering all $J_a$'s equal to $J$. It is worth pointing out the difference with the equilibrium 
situation at $T=0$. There, the properties of the system are obtained from the ground state, \textit{i.e.} the solution 
with minimal action (energy). The ground state can be selected through the introduction of a Boltzmann-like 
weighting factor with an auxiliary temperature in the limit where the latter is taken to zero.\cite{tissier11,tissier12b} 
The selection of the extremal states is thus quite different.

To proceed further, as explained in detail in Ref.~[\onlinecite{tissier11}], the above functional can be reexpressed with the help of 
auxiliary fields through standard field-theoretical techniques.\cite{zinnjustin89,parisi79} This leads to a ``superfield theory'' with a large group 
of symmetries and supersymmetries.\cite{tissier11}

The next step consists in applying an exact RG formalism to this superfield theory.  This can be done by progressively including 
the contribution of the fluctuations of the superfield on longer length scales, or alternatively with shorter 
momenta.\cite{wilson74} Technically, this can be implemented through the addition to the bare action of an 
``infrared (IR) regulator'' depending on a running IR scale $k$; its role is to suppress, in the generating functional derived from 
Eq.~(\ref{eq_generating_func1}), the integration over modes with momentum $\vert q \vert \lesssim k$.\cite{wetterich93,berges02,tarjus04}

The central quantity of our RG approach is the $k$-dependent ``effective average action'',\cite{wetterich93,berges02} $\Gamma_k$. 
This functional exactly interpolates between the bare action at the 
microscopic (or UV) scale $k=\Lambda$, which then corresponds to the mean-field approximation where no fluctuations are accounted for, 
and the exact effective action (Gibbs free energy) when $k=0$. The latter is the generating functional of the so-called ``1-particle irreducible'' 
(1PI) correlation functions\cite{zinnjustin89} and its knowledge entails a full description of the statistical properties of the extremal states, hence of 
the out-of-equilibrium hysteresis behavior of the RFIM. Expanding $\Gamma_k$ in increasing number of unrestricted sums over 
copies (or replicas) generates a cumulant expansion for the renormalized disorder at the scale $k$.\cite{tissier11}

The RG flow of $\Gamma_k$ is generated by continuously decreasing the IR scale $k$. This leads to an 
\textit{exact functional RG equation},\cite{wetterich93,berges02} from which one  derives an exact 
hierarchy of coupled functional RG equations for the cumulants of the renormalized disorder.

An important simplification occurs in the situation of interest here. As already mentioned, the hysteresis loop corresponds to the limit of 
infinite auxiliary source, $\hat J \rightarrow \pm \infty$, or in the Legendre transformed setting, the limit of infinite 
auxiliary field, $\hat \phi \rightarrow \pm \infty$ ($\hat \phi_a$ is the average of the auxiliary field introduced as conjugate of the source $J_a$ 
to reexpress Eq. (\ref{eq_generating_func1})\cite{tissier11}). The main point is that \textit{the uniqueness of the extremal states} 
(for each branch separately) translates in the present superfield framework 
into the formal property of the random generating functional which we called  
``Grassmannian ultralocality"\cite{tissier11} and which greatly simplifies the formalism. A discussion of this property and technical details are 
provided in the Supplemental Material.\ref{supplemental}

After some  algebra, we end up with exact RG functional equations for the cumulants of the 
renormalized disorder, $\Gamma_{k1}[\phi_a]$, $\Gamma_{k2}[\phi_a,\phi_b]$, etc, or more precisely for the cumulants of the 
renormalized random field, $\Gamma_{k1;x_1}^{(1)}[\phi_a]$, $\Gamma_{k2;x_1,x_2}^{(1,1)}[\phi_a,\phi_b]$, etc,  with 
the \textit{physical} fields $\phi_a$ only as arguments (superscripts denote functional differentiation with respect to the arguments). 
As an illustration, the equation for the first cumulant reads
\begin{equation}
\label{eq_flow_Gamma1(1)_UL}
\begin{split}
\partial_t\Gamma_{k1;x}^{(1)}\left[\phi_1\right ]=& -\dfrac{1}{2} \tilde{\partial}_t \frac{\delta}{\delta \phi_1(x)} \int_{x_2}\int_{x_3} 
\left(\Gamma _{k1}^{(2)}[ \phi_1]+\widehat R_k\right) ^{-1}_{x_2x_3}\\&
\times \bigg(\Gamma_{k2;x_2,x_3}^{(11)}\left[\phi_1,\phi_1\right ] - \widetilde{R}_{k;x_2,x_3}\bigg) \, .
\end{split}
\end{equation}
where $t=\log(k/\Lambda)$, $\widehat{R}_k(q^2)$ and $\widetilde{R}_k(q^2)$ are IR regulators: 
$\widehat{R}_k(q^2)$ gives a mass $\sim k^2$ for modes with $\vert q\vert \lesssim k$ and is essentially zero for modes with 
$\vert q\vert \gtrsim k$, while $\widetilde{R}_k(q^2)$ suppresses fluctuations of the random field and is related to $\widehat{R}_k(q^2)$ in   
a way that is  compatible with the underlying supersymmetry of the theory.\cite{tissier11,tissier12b} Finally, $\widetilde{\partial}_t$ is a short-hand 
notation indicating a derivative with respect to $t$ that acts on the cutoff functions only (\textit{i.e.},  $\widetilde{\partial}_t 
\equiv \partial_t \widehat{R}_k\, \delta/\delta \widehat{R}_k + \partial_t \widetilde{R}_k \, \delta/\delta \widetilde{R}_k$). The auxiliary fields have 
thus completely dropped out of the equations.

One finds that Eq.~(\ref{eq_flow_Gamma1(1)_UL}) coincides with the derivative with respect to $\phi_1(x)$ of  the exact RG flow equation 
followed by $\Gamma_{k1}\left[\phi_1\right ]$ for the RFIM at equilibrium (see Eq.~(7) of Ref.~[\onlinecite{tissier12b}]). It is easily derived that this 
generalizes to all higher-order cumulants, so that the exact hierarchies of RG flow equations for the cumulants of the renormalized 
random field $\Gamma_{kp;x_1,\cdots,x_p}^{(1 \cdots 1)}\left[\phi_1,\cdots, \phi_p\right ]$ for the RFIM in and out of equilibrium 
are identical. This is a central result of the present work, and the main ingredient is the uniqueness of the selected 
stationary states at zero temperature.\cite{footnote}

We have therefore shown that \textit{the out-of-equilibrium hysteresis behavior and the ground-state physics are described by the same 
exact RG equations}. For the equilibrium case, one knows that there exists a fixed point associated with critical behavior. This fixed 
point has a $Z_2$ symmetry, \textit{i.e.} all functions are symmetric under the inversion of the fields 
(local magnetizations). It has two relevant directions, one corresponding to a $Z_2$ symmetric perturbation and associated with the disorder 
strength that must be fine-tuned to be exactly at criticality and the other being non $Z_2$ symmetric and associated with the external  
source (which in some sense is also tuned to be zero, which amounts to staying in the $Z_2$ symmetric subspace). 

The out-of-equilibrium critical point on the other hand has no $Z_2$ symmetry: it takes place at nontrivial values 
of the external source (magnetic field) and of the field (magnetization):  $J_c>0$,  $\phi_c \neq 0$ for the ascending 
branch of the hysteresis loop and, due to the statistical $Z_2$ symmetry, $- J_c$,  
$- \phi_c$ for the descending  branch (see Fig.~1). This implies that the initial condition to the exact RG flow equations, \textit{i.e.} 
the mean-field description at the microscopic scale, has no $Z_2$ symmetry. (This is akin to the situation encountered 
when relating the liquid-gas critical point of a genuine fluid that has no particle-hole symmetry to that of the simple Ising model with $Z_2$ 
symmetry.)

To show that non $Z_2$ symmetric initial conditions appropriate for describing out-of-equilibrium criticality can flow 
under RG transformation to the already characterized $Z_2$ symmetric equilibrium fixed point, we consider the 
nonperturbative approximation scheme for the effective average action that we have already introduced 
in our previous work on the RFIM at equilibrium.\cite{tissier11,tissier12b} It combines a truncation in the ``derivative expansion'', \textit{i.e.} 
an expansion in the number of spatial derivatives of the fundamental fields for approximating the long-distance behavior of the 1PI correlation 
functions, and a truncation in the expansion in cumulants of the renormalized disorder. The scheme also ensures that the symmetries and 
supersymmetries of the theory are not explicitly violated, which turns out to be an important issue for a proper description of  
``dimensional reduction'' and its breakdown.\cite{tissier11,tissier12b}  The approximation scheme then leads to a closed set of coupled 
nonperturbative functional RG equations that can be solved numerically.

When formulated at the level of the cumulants of the renormalized random field, the ansatz takes the form
\begin{equation}
\begin{aligned}
\label{eq_ansatz}
&\Gamma_{k1;x_1}^{(1)}[\phi]= U'_k(\phi(x_1))+\frac{\delta}{\delta \phi(x_1)}\big \{\frac{1}{2}Z_k(\phi(x_1))[\partial \phi(x_1)]^2
\big \} \\&
\Gamma_{k2;x_1,x_2}^{(11)}[\phi_1,\phi_2]=\Delta_k(\phi_1(x_1),\phi_2(x_2)),
\end{aligned}
\end{equation}
with the higher-order cumulants set to zero. For concreteness, we focus on the critical point along the ascending branch 
of the hysteresis loop, with $J_c >0$.
After insertion in the hierarchy of exact RG equations [Eq.~(\ref{eq_flow_Gamma1(1)_UL}) and its higher-order counterparts], 
the above ansatz provides three coupled flow equations for $U'_k(\phi)$, which describes the renormalized source 
as a function of magnetization, the so-called ``field-renormalization'' function $Z_k(\phi)$, and the second cumulant of the renormalized 
random field $\Delta_k(\phi_1,\phi_2)$.

These flow equations are supplemented by an initial condition at the microscopic (UV) scale $k=\Lambda$. It corresponds to a mean-field 
approximation where only some  
coarse-graining over short-ranged fluctuations has been carried out (see {\it e.g.}  [\onlinecite{mlr10}]). 
The crucial point is that the bare action has no $Z_2$ 
symmetry around the out-of-equilibrium critical point. The initial condition can then be taken with the same form as 
in Eq.~(\ref{eq_ansatz}) with $Z_{\Lambda}=1$,  $\Delta_{\Lambda}=\Delta_B$, 
and $U'_{\Lambda}(\phi)$ generically given by
$U'_{\Lambda}(\phi)=J_{\Lambda}+r_{\Lambda}\phi+\frac{\lambda_{\Lambda}}{2}\phi^2+\frac{u_{\Lambda}}{3!}\phi^3
+\frac{v_{\Lambda}}{4!}\phi^4$,
where higher-order terms can be dropped as they do not change the universal properties.

To cast the RG flow equations in a dimensionless form that allows one to investigate the critical physics at long length scales, one 
must introduce scaling dimensions. This is the second operation of any RG transformation. Near a zero-temperature  
fixed point, the renormalized temperature is irrelevant and is characterized by an exponent $\theta>0$.\cite{villain84,fisher86}  One then has the 
following scaling dimensions: $Z_{k} \sim k^{-\eta}$, $\phi-\phi_c  \sim k^{(d-4+\bar \eta)/2}$, 
$U'_k-J_c\sim k^{(d-2\eta+\bar \eta)/2}$, $\Delta_k \sim k^{-(2\eta- \bar \eta)}$,
where $\phi_c$ and $J_c$ respectively denote the  values of the magnetization and the magnetic field at the 
out-of-equilibrium critical point (see above), and the exponents $\theta$, $\eta$ and $\bar\eta$ are related 
through $\theta=2+\eta-\bar\eta$.

Due to the lack of $Z_2$ inversion symmetry, two relevant parameters must be fine-tuned to reach the critical point. In practice, we 
account for the additional condition by defining a displaced field variable $\tilde\phi=\phi-\phi_{r,k}$ where $\phi_{r,k}$ is fixed such that 
the third derivative of the renormalized potential is zero all along the flow: $U'''_k(\phi_{r,k})=0$. If indeed the critical system flows to a fixed 
point where $Z_2$ symmetry is restored, then $\phi_{r,k}$ flows to the critical value $\phi_c$ and $U'_k(\phi_{r,k})$ flows to $J_c$ 
in the limit $k\rightarrow 0$.

Using lower-case letters, $u''_k, z_k,\delta _k,  \tilde\varphi$, to denote the dimensionless counterparts of 
$U''_k, Z_k,\Delta _k,  \tilde\phi$,  the dimensionless form of the flow equations can be symbolically written as
\begin{equation}
\label{eq_flow_dimensionless}
\begin{split}
&\partial_t u''_k(\tilde\varphi)=\beta_{u'',k}(\tilde\varphi),\; 
\partial_t z_k(\tilde\varphi)=\beta_{z,k}(\tilde\varphi),\\&
\partial_t \delta_k(\tilde\varphi_1,\tilde\varphi_2)=\beta_{\delta,k}(\tilde\varphi_1,\tilde\varphi_2)\, ,
\end{split}
\end{equation}
where the beta functions in the right-hand sides themselves depend on $u''_k$, $z_k$, $\delta_k$ and their derivatives. As already 
stressed above, these flow equations are \textit{the same} as for the RFIM at equilibrium; they are given in Ref.~[\onlinecite{tissier12b}] 
and not reproduced here.
\begin{figure}[ht]
\includegraphics[width=\linewidth]{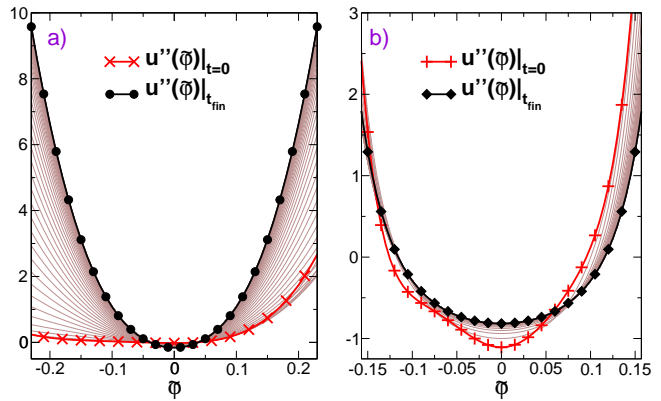}
\caption{
Nonperturbative RG evolution of the dimensionless ``mass" function $u''_k(\tilde\varphi)$ for $d=5.5 >d_{DR}\simeq 5.1$ (a) and 
$d=4<d_{DR}$ (b). The initial condition 
(in red) is asymmetric, but the asymmetry gradually decreases along the flow and vanishes at the 
fixed point (black curve). Furthermore, the fixed-point function for $u''_k(\tilde\varphi)$ is identical to that for the equilibrium critical point. }
\end{figure}
The nonperturbative RG equations can be solved for any spatial dimension $d$ and a variety of initial conditions (yet two parameters must 
be fine-tuned to reach the critical fixed point). In all cases, we find that the flow leads to the $Z_2$ symmetric fixed point already derived for 
the equilibrium critical point. We illustrate the outcome for two cases (see Fig. 2): one is above the critical dimension for 
dimensional-reduction breakdown, $d_{DR}\simeq 5.1$,\cite{tissier11,tissier12b} and is therefore exactly described by 
the $d \rightarrow d-2$ dimensional-reduction property ; the other is below $d_{DR}$ and does not follow dimensional reduction. 
In both situations, one can clearly see that the asymmetry of $u''_k(\tilde\varphi)$ 
eventually decreases and vanishes when reaching the fixed point. (The same is observed for the other functions $z_k(\tilde\varphi)$ and 
$\delta _k(\tilde\varphi,\tilde\varphi)$ but is not displayed here.) \textit{The $Z_2$ symmetry is thus asymptotically 
restored and the fixed point \textit{exactly} coincides with that found for the equilibrium criticality.}

The conclusion is that the critical behaviors of the RFIM in and out of equilibrium are in the same universality class, with 
the same critical exponents, the same scaling functions and the same avalanche-size distribution. This gives a solid 
theoretical foundation to the empirical numerical findings. 
Along the way, the above developments also prove that the in- and 
out-of-equilibrium critical behaviors of fluids in a disordered porous material, which are both described by non $Z_2$ 
symmetric theories,\cite{tarjusHRT} are in this same universality class. Our present work therefore unifies a very large class of 
collective phenomena in and out of equilibrium that involve interactions and disorder.

\appendix 

\section{Supplemental Material}
\label{supplemental}

\subsection{Grassmannian ultralocality and exact functional RG equations}

As explained in detail in Ref.~[\onlinecite{tissier11}], the functional in Eq.~(3) of the main text can be reexpressed through standard 
field-theoretical techniques\cite{zinnjustin89,parisi79} as that of a ``superfield theory'' with a large group of symmetries and supersymmetries. 
In a nutshell, one introduces auxiliary (bosonic) fields $\hat{\varphi}_a$ to ``exponentiate'' the delta functional, pairs of auxiliary (fermionic) 
fields $\psi_a,\bar{\psi}_a$ to ``exponentiate'' the determinant and one averages over the Gaussian random field.\cite{parisi79} By a 
Legendre transform, one then obtains the ``effective action'' (Gibbs free energy), $\Gamma[\{\Phi_a\}]$ where the $\Phi_a$'s are now 
``superfields'' leaving in a ``superspace'' spanned by the $d$-dimensional Euclidean coordinate $x$ and two anti-commuting 
Grassmann coordinates $\theta, \bar\theta$:\cite{parisi79,zinnjustin89,tissier11}
\begin{equation}
\label{eq_superfield}
\Phi_a(x,\theta, \bar\theta)=\phi_a(x) + \bar\theta \Psi_a(x) + \bar\Psi_a(x) \theta +\bar \theta \theta \hat \phi_a(x)\,,
\end{equation}
where $\phi_a(x), \Psi_a(x), \bar\Psi_a(x), \hat \phi_a(x)$ denote the averages of the physical field and of the associated auxiliary 
fields in copy $a$. 

After having introduced infrared (IR) regulators, one may define an effective averaged action $\Gamma_k[\{\Phi_a\}]$ which is the effective 
action of the system at the scale $k$.\cite{wetterich93,berges02} Its expansion in increasing number of unrestricted sums over copies generates 
(modulo some inessential 
subtleties\cite{tissier11}) the  cumulant expansion for the renormalized disorder: 
\begin{equation}
\label{eq_free_replica_sums}
\Gamma_k\left[\{ \Phi_a\}\right ]=
\sum_a \mathsf\Gamma_{k1}[\Phi_a]-\frac{1}{2}\sum_{a,b}\mathsf\Gamma_{k2}[\Phi_a,\Phi_b] + \cdots
\end{equation}
where $\mathsf\Gamma_{p}$ is essentially the $p$th cumulant of the renormalized disorder.\cite{tissier11} 
Such an expansion in increasing number of free sums over copies lead to systematic algebraic manipulations that we have repeatedly used.

As recalled in the main text, the evolution of $\Gamma_k[\{\Phi_a\}]$ with $k$ is described by an exact functional 
renormalization-group (RG) equation, from which one can derive a hierarchy of exact functional RG equations for the cumulants. 
For the sake of illustration, we give below the exact RG equation for the first cumulant:
\begin{equation}
\label{eq_ERGE_Gamma1}
\begin{split}
\partial_t \mathsf \Gamma_{k1}[\Phi_1]=&\frac 12  \tilde{\partial}_t \tr  \Big \{ \log \Big (\mathsf \Gamma_{k1}[\Phi_1] 
+ \widehat R_{k} I\Big ) + \Big(\mathsf \Gamma _{k1}^{(2)}[ \Phi_1]\\& + \widehat R_k I \Big) ^{-1}
\Big (\mathsf\Gamma_{k2}^{(11)}[\Phi_1,\Phi_1 ] - \widetilde{R}_{k} I\Big)\Big \} ,
\end{split}
\end{equation}
where $t=\log(k/\Lambda)$, the trace involves summing over copy indices and integrating over superspace, and superscripts denote  
functional differentiation with respect to the superfield arguments; $I$ is the identity, $I_{(a,\theta_1,\bar\theta_1)(b,\theta_2,\bar\theta_2)}
=\delta_{ab}\delta_{\bar\theta_1,\bar\theta_2}\delta_{\theta_1,\theta_2}$, where $\delta_{ab}$ is the Kronecker symbol and due to the 
anticommuting properties of the Grassmann variables,\cite{zinnjustin89} $\delta_{\bar\theta_1,\bar\theta_2}\delta_{\theta_1,\theta_2}=
(\bar\theta_1-\bar\theta_2)(\theta_1-\theta_2)$; $\widehat{R}_k(q^2)$ and $\widetilde{R}_k(q^2)$ are IR cutoff functions and 
$\widetilde{\partial}_t$ is a short-hand notation indicating a derivative with respect to $t$ that acts on the cutoff functions only (see the main text). 

The above RG equation, and the whole hierarchy for higher-order cumulants, is exact but too formal to be useful as such. A major 
simplification however occurs when the generating functional is built from a unique stationary state (in each replica), which is expected  here 
in the limit of infinite auxiliary source , $\hat J \rightarrow \pm \infty$, or in the Legendre transformed setting, the limit of infinite 
auxiliary field, $\hat \phi \rightarrow \pm \infty$. The uniqueness of the extremal states indeed  
translates in the present superfield framework in the formal property of the random generating functional that we called  
``Grassmannian ultralocality".\cite{tissier11} The cumulants are then ``ultralocal", \textit{i.e.}
\begin{equation}
\begin{aligned}
&\label{eq_app_ULcumulants}
\mathsf\Gamma_{k1}[\Phi_1]=\int_{\underline{\theta_1}}\Gamma_{k1}[\Phi_1(\underline{\theta_1})] \\&
\mathsf\Gamma_{k2}[\Phi_1,\Phi_2]=\int_{\underline{\theta_1}}\int_{\underline{\theta_2}}
\Gamma_{k2}[\Phi_1(\underline{\theta_1}),\Phi_2(\underline{\theta_2})]\, ,
\end{aligned}
\end{equation}
etc, where we have grouped the two Grassmann coordinates $\theta,\bar\theta$ in the notation $\underline{\theta}$ and 
$\int_{\underline{\theta}} \equiv \int \int d\theta d\bar\theta$. $\Gamma_{k1}$ $\Gamma_{k2}$, etc, in the right-hand sides
only depends on the superfields at the explicitly displayed ``local" Grassmann coordinates, hence the name ``Grassmannian ultralocality". 
(On the other hand, the dependence on the Euclidean coordinates, which is left implicit, is \textit{not} purely local). 

The property of Grassmannian ultralocality is also true for the equilibrium case, where the generating functional is dominated by the 
ground state which is also unique for a given sample (except, again, for a set of conditions of measure zero); it then greatly simplifies the 
exact functional RG equations.\cite{tissier11,tissier12b}  In the present case, one must however proceed differently. We first differentiate 
the exact RG equations, such as Eq.~(\ref{eq_ERGE_Gamma1}),  in order to obtain RG equations for the cumulants of the 
\textit{renormalized random field}, $\mathsf\Gamma_{k1;x_1\underline{\theta_1}}^{(1)}[\Phi_1]$, 
$\mathsf\Gamma_{k2;x_1\underline{\theta_1};x_2\underline{\theta_2}}^{(11)}[\Phi_1,\Phi_2]$, etc. We next evaluate the latter equations 
at the (external) Grassmann coordinates $\theta_a=\bar\theta_a=0$, for $a=1,2,\cdots$. After some straightforward algebra, we end up with   
exact RG functional equations for the cumulants $\Gamma_{kp;x_1,\cdots,x_p}^{(1 \cdots 1)}$ with the \textit{physical} fields $\phi_a$ only 
as arguments. For instance, the equation for the first cumulant is given in Eq.~(4) of the main text. These equations exactly coincide with those 
obtained for the same quantities, after using there the very same property of Grassmannian ultralocality, in the equilibrium case.\cite{tissier11}

\subsection{Non-ultralocal corrections}

In the above derivation, we have actually used a short-cut that needs justification. Indeed, we have taken the limit $\hat \phi \rightarrow 
\infty$ before a full account of the fluctuations and the limit $k\rightarrow 0$. The correct procedure is instead to solve the exact 
RG flow down to $k=0$ for $\hat \phi$ very large but finite and then take $\hat \phi$ to infinity. For large but finite $\hat \phi$, 
there are corrections to the Grassmannian ultralocality. It can however be checked that these corrections become irrelevant as one 
approaches the fixed point when $k\rightarrow 0$ and therefore give rise to only subdominant contributions. This is what we discuss now. 

We now illustrate the structure of the functional RG flow in the presence of  ``non-ultralocal'' components by looking at the corrections 
in the first cumulant and assuming that all other cumulants are purely ``ultralocal'' in both Grassmann and Euclidean coordinates.
More specifically, we consider 
\begin{equation}
\begin{aligned}
\label{eq_cumulant_1_nonultralocal_correction}          
&\mathsf \Gamma_{k1}[\Phi_{1}] = \int_{\underline{\theta}_1} 
\Big (\Gamma_{k1}^{UL}[ \Phi_1(\underline{\theta}_1)] + \\&\Gamma_{k1}^{NUL}[ \Phi_1(\underline{\theta}_1),
\partial_{ \theta_1} \Phi_1(\underline{\theta}_1),\partial_{\bar \theta_1} \Phi_1(\underline{\theta}_1), 
\partial_{ \theta_1}\partial_{\bar \theta_1} \Phi_1(\underline{\theta}_1)] \Big )
\end{aligned}
\end{equation}
where $\Gamma_{k1}^{NUL}$ is non-ultralocal in the Grassmann coordinates (\textit{i.e.} depends on the derivatives) 
but ultralocal in the Euclidean coordinates, and for $p \geq 2$,
\begin{equation}
\begin{aligned}
\label{eq_cumulant_p_ultralocal}          
&\mathsf \Gamma_{kp}[\Phi_{1},...,\Phi_{p}] =\\&\int_{\underline{\theta}_1}...\int_{\underline{\theta}_p}
\int_{x_1}...\int_{x_p} \Gamma_{kp}^{UL}[ \Phi_1(x_1,\underline{\theta}_1),..., \Phi_p(x_p,\underline{\theta}_p)]\,.
\end{aligned}
\end{equation}
 
By virtue of the supersymmetries of the theory, the non-ultralocal part of the first cumulant can be rewritten in terms of 
components in the following form:
\begin{equation}
\begin{aligned}
\label{eq_cumulant_1_nonultralocal_components}          
\mathsf \Gamma_{k1}^{NUL}[\Phi_{1}] =&  \int_x
\Big [\hat \phi_1(x)\, \gamma_{k1}( \phi_1(x), \hat \phi_1(x))\\&
+ \Psi_1(x)\bar \Psi_1(x)\, \gamma_{k1}^{(10)}( \phi_1(x), \hat \phi_1(x))\Big ] \,.
\end{aligned}
\end{equation}
In principle, all manipulations should involve the fermionic fields  $\Psi_1, \bar \Psi_1$, but it turns out that supersymmetries again lead to 
simplifications and that the same results are obtained by setting these fields to zero, which we do here to simplify the presentation.

The second functional derivative of the effective average action $\Gamma_{k;(a\underline{\theta}_1)(a \underline{\theta}_2)}^{(2)}$ 
that enters in the functional RG equations can be decomposed as\cite{tissier11}
\begin{equation}
\label{eq_app_free_replica_sums}
\mathsf \Gamma_{k;(a\underline{\theta}_1)(b \underline{\theta}_2)}^{(2)}=
\widehat {\mathsf\Gamma}_{k;a\underline{\theta}_1 \underline{\theta}_2}^{(2)}\delta_{ab}+
\widetilde {\mathsf\Gamma}_{k;(a\underline{\theta}_1)(b \underline{\theta}_2)}^{(2)}
\end{equation}

After adding the IR regulators, the ``hat'' and ``tilde'' components have the following general structure:
\begin{equation}
\begin{split}
\label{eq_hatGamma2_NUL}
\widehat {\mathsf\Gamma}_{k;a\underline{\theta}_1 \underline{\theta}_2}^{(2)}&[ \{ \phi_a,\hat \phi_a\}]
+\widehat R_k \delta_{\underline{\theta}_1 \underline{\theta}_2}
=\widehat {A}_{k;a} +\bar \theta_1 \theta_1\bar \theta_2 \theta_2\widehat {B}_{k;a}\\&
+(\bar \theta_1 \theta_1+\bar \theta_2 \theta_2)
\widehat {C}_{k;a}-(\bar \theta_1 \theta_2+\bar \theta_2 \theta_1) \widehat {E}_{k;a}
\end{split}
\end{equation}
\begin{equation}
\begin{aligned}
\label{eq_tildeGamma2_NUL}
\widetilde {\mathsf\Gamma}_{k;(a\underline{\theta}_1)(b \underline{\theta}_2)}^{(2)}[ \{\phi_a,\hat \phi_a\}]+
\widetilde R_k\delta_{\underline{\theta}_1 \underline{\theta}_2}&=
\widetilde {A}_{k;ab} +\bar \theta_1 \theta_1\bar \theta_2 \theta_2\widetilde {B}_{k;ab}
\\&+\bar \theta_1 \theta_1\widetilde {C}_{k;ab}
+\bar \theta_2 \theta_2\widetilde {D}_{k;ab} \, ,
\end{aligned}
\end{equation}
and to the lowest order of the expansions in increasing number of free sums over copies\cite{tissier11}
(leaving implicit the dependence on the Euclidean coordinates):
\begin{equation}
\begin{aligned}
\label{eq_expand_Gamma_2hatNUL}
&\widehat{A}_{k;a}[\phi_a,\hat \phi_a] =2 \gamma_{k1}^{(01)}[\phi_{a},\hat \phi_a]+ 
\hat \phi_a\gamma_{k1}^{(02)}[\phi_{a},\hat \phi_a]\\&
\widehat{B}_{k;a}[\phi_a,\hat \phi_a] =\hat \phi_a \widehat\Gamma_{k1}^{UL(3)}[\phi_{a}]\\&
\widehat{C}_{k;a}[\phi_a,\hat \phi_a] =\\& \widehat\Gamma_{k1}^{UL(2)}[\phi_{a}]+\widehat R_k
+\gamma_{k1}^{(01)}[\phi_{a},\hat \phi_a]+\hat \phi_a\gamma_{k1}^{(11)}[\phi_{a},\hat \phi_a]\\&
\widehat{E}_{k;a}\phi_a,\hat \phi_a] =\widehat\Gamma_{k1}^{UL(2)}[\phi_{a}]+\widehat R_k 
+\gamma_{k1}^{(10)}[\phi_{a},\hat \phi_a]
\end{aligned}
\end{equation}
and
\begin{equation}
\begin{aligned}
\label{eq_expand_Gamma_2tildeNUL}
&\widetilde{A}_{k;ab}[\phi_a,\hat \phi_a,\phi_b,\hat \phi_b]=-\Gamma_{k2}^{UL(11)}[\phi_{a},\phi_{b}]+\widetilde R_k\\& 
\widetilde{B}_{k;ab}[\phi_a,\hat \phi_a,\phi_b,\hat \phi_b]=-\Gamma_{k2}^{UL(22)}[\phi_{a},\phi_{b}]\\&
\widetilde{C}_{k;ab}[\phi_a,\hat \phi_a,\phi_b,\hat \phi_b]=-\hat \phi_a\Gamma_{k2}^{UL(21)}[\phi_{a},\phi_{b}]\\&
\widetilde{D}_{k;ab}[\phi_a,\hat \phi_a,\phi_b,\hat \phi_b]=-\hat \phi_b\Gamma_{k2}^{UL(12)}[\phi_{a},\phi_{b}]\;.
\end{aligned}
\end{equation}

The full propagator $\mathcal P_{k;(a_1\underline{\theta}_1)(a_2 \underline{\theta}_2)}$, which is the inverse 
of $\mathsf \Gamma_k^{(2)} +\mathcal R_k$ (where $\mathcal R_k$ collects the two IR regulators), has the same structure 
as in Eqs.~(\ref{eq_app_free_replica_sums},\ref{eq_hatGamma2_NUL},\ref{eq_tildeGamma2_NUL}) 
with
\begin{equation}
\begin{split}
\label{eq_hatP0_NUL}
&\widehat {\mathcal P}_{k;a\underline{\theta}_1 \underline{\theta}_2}[ \{ \phi_a,\hat \phi_a\}] 
=\\&\widehat {Q}_{k;a} +\bar \theta_1 \theta_1\bar \theta_2 \theta_2\widehat {S}_{k;a}+(\bar \theta_1 \theta_1+\bar \theta_2 \theta_2)
\widehat {P}_{k;a}-(\bar \theta_1 \theta_2+\bar \theta_2 \theta_1) \widehat {T}_{k;a}
\end{split}
\end{equation}
and an expression similar to Eq.~(\ref{eq_tildeGamma2_NUL}) for 
$\widetilde {\mathcal P}_{k;(a\underline{\theta}_1)(b \underline{\theta}_2)}[ \{\phi_a,\hat \phi_a\}]$.

At the lowest order of the expansion in increasing free sums over copies, the components of $\widehat {\mathcal P}_{k}$ 
and $\widehat {\mathsf\Gamma}_{k}^{(2)}$ are related by 
\begin{equation}
\begin{aligned}
\label{eq_hat_inversion}
&\widehat{Q}_{k}[\phi_a,\hat \phi_a] =- \big (\widehat{C}_{k}\widehat{C}_{k}
-\widehat{A}_{k}\widehat{B}_{k}\big )^{-1}\widehat{A}_{k}\\&
\widehat{S}_{k}[\phi_a,\hat \phi_a] =-\big (\widehat{C}_{k}\widehat{C}_{k}
-\widehat{A}_{k}\widehat{B}_{k}\big)^{-1}\widehat{B}_{k}\\&
\widehat{P}_{k}[\phi_a,\hat \phi_a] =\big (\widehat{C}_{k}\widehat{C}_{k}
-\widehat{A}_{k}\widehat{B}_{k}\big)^{-1}\widehat{C}_{k}\\&
\widehat{T}_{k}[\phi_a,\hat \phi_a] =\widehat{E}_{k}^{-1}\,,
\end{aligned}
\end{equation}
where one should keep in mind that the components are operators in Euclidean space. 

On the other hand, the ``tilde'' components of the propagator are obtained at the lowest order of the expansion in free sums over copies 
from 
\begin{equation}
\begin{split}
\label{eq_tildeP0_inversion}
&\widetilde {\mathcal P}_{k;(a\underline{\theta}_1)(b \underline{\theta}_2)}[\phi_a,\hat \phi_a,\phi_b,\hat \phi_b]
=- \int_{\underline{\theta}_3}\int_{\underline{\theta}_4}\widehat {\mathcal P}_{k;\underline{\theta}_1\underline{\theta}_3}[ \phi_a,\hat \phi_a]
\\&\times \Big (\widetilde {\mathsf\Gamma}_{k;\underline{\theta}_3 \underline{\theta}_4}^{(2)}[\phi_a,\hat \phi_a,\phi_b,\hat \phi_b ]
+\widetilde R_k\Big )
\widehat {\mathcal P}_{k;\underline{\theta}_4\underline{\theta}_2}[ \phi_b,\hat \phi_b]\, .
\end{split}
\end{equation}
The algebraic manipulations are straightforward but the resulting expressions are too lengthy to be reproduced here. We stress that \textit{no} 
approximations are involved in deriving results at the lowest order of the expansion in free sums over copies. The higher orders are not needed.

\subsection{Functional RG equations with non-ultralocal corrections}

We can now collect the above results and insert them in the exact RG equation for the first cumulant, Eq.~(\ref{eq_ERGE_Gamma1}). 
This leads to
\begin{equation}
\begin{aligned}
\label{eq_flow_cum1_1}
&\partial_t\mathsf \Gamma_{k1}[\Phi_1]=\\& \frac{1}{2}\int_{x_1x_2}\int_{\underline{\theta}_1,\underline{\theta}_2}
\Big \{\partial_t\widetilde{R}_k(x_1-x_2)\widehat {\mathcal P}_{k;(x_1,\underline{\theta}_1)(x_2, \underline{\theta}_2)}[ \phi_1,\hat \phi_1]
\\&
+ \partial_t\widehat{R}_k(x_1-x_2)\, \delta_{\underline{\theta}_1,\underline{\theta}_2}
\Big (\widehat {\mathcal P}_{k;(x_1,\underline{\theta}_1)(x_2, \underline{\theta}_2)}[ \phi_1,\hat \phi_1] 
+\\&\widetilde {\mathcal P}_{k;(x_1,\underline{\theta}_1)(x_2, \underline{\theta}_2)}[\phi_1,\hat \phi_1,\phi_2,\hat \phi_2] \Big ) \Big \}\,.
\end{aligned}
\end{equation}
After taking a functional derivative with respect to $\Phi_1(x,\underline{\theta})$, evaluating the outcome for $\theta=\bar\theta=0$ 
and using Eqs.~(\ref{eq_cumulant_1_nonultralocal_correction},\ref{eq_cumulant_p_ultralocal}, 
\ref{eq_cumulant_1_nonultralocal_components}), one obtains an explicit 
RG flow equation for $\Gamma_{k1;x}^{(1)UL}[\phi_{1}] + \gamma_{k1}( \phi_1(x), \hat \phi_1(x)) + \hat \phi_1(x) 
\gamma_{k1}^{(01)}( \phi_1(x), \hat \phi_1(x))$. To keep the presentation in a reasonable format, we further evaluate the equation for 
spatially uniform fields $ \phi_1(x)\equiv \phi_1, \hat \phi_1(x)\equiv \hat \phi_1$, so that it simplifies to
\begin{equation}
\begin{aligned}
\label{eq_flow_cum1_2}
&\partial_tU_{k}^{(1)}(\phi_{1}) + \partial_t \big [\gamma_{k1}( \phi_1, \hat \phi_1) + \hat \phi_1 
\gamma_{k1}^{(01)}( \phi_1, \hat \phi_1)\big ]=\\& \frac{1}{2}\frac{\delta}{\delta \hat\phi_1}\int_{q}
\Big \{\partial_t\widetilde{R}_k(q^2)
\widehat S_{k}(q^2) + \partial_t\widehat{R}_k(q^2) \Big (2\big [\widehat P_{k}(q^2) \\&- 
\widehat T_{k}(q^2)\big ]  + 2 \widehat S_{k}(q^2) \big [ \Gamma_{k2}^{(11)}(q^2;\phi_1,\phi_1) -
\tilde{R}_k(q^2)\big ] \widehat P_{k}(q^2) \\&
+2 \widehat Q_{k}(q^2) \hat{\phi}_1^2\Gamma_{k2}^{(22)}(q^2;\phi_1,\phi_1) \widehat P_{k}(q^2) + 
\big [\widehat Q_{k}(q^2)\widehat S_{k}(q^2)  \\& 
+ \widehat P_{k}(q^2)^2\big ] \hat{\phi}_1\big [ \Gamma_{k2}^{(21)}(q^2;\phi_1,\phi_1)
+\Gamma_{k2}^{(12)}(q^2;\phi_1,\phi_1) \big ]\Big )\Big \}.
\end{aligned}
\end{equation}
where $U_{k}(\phi_1)$ is the effective average potential, \textit{i.e.} the component of the first cumulant that is ultralocal in both  
Euclidean and Grassmann coordinates; $\widehat P_{k}(q^2),\widehat Q_{k}(q^2),\widehat S_{k}(q^2),\widehat T_{k}(q^2)$ are functions of 
$\phi_1$ and $\hat\phi_1$.

Since we are interested in showing that the non-ultralocal corrections give subdominant corrections near the fixed point the limit 
$\vert \hat \phi \vert \rightarrow +\infty$, it is sufficient to consider an expansion in $1/\vert \hat \phi\vert$. For convenience we choose to study the 
descending branch of the hysteresis loop with $\hat J>0$ and $\hat \phi>0$. The non-ultralocal component $\gamma_{k1}(\phi,\hat\phi)$ 
has an expansion of the form
\begin{equation}
\label{eq_expansion_large}
\gamma_{k1}(\phi,\hat{\phi})=\frac{1}{\hat{\phi}^p}\Big(X_{k0}(\phi)+\frac{X_{k1}(\phi)}{\hat{\phi}}+\cdots\Big).
\end{equation}
with $p>1$. 

It is easily realized that when the above expansion is inserted in the functional RG equation, Eq.~(\ref{eq_flow_cum1_2}), the right-hand 
side can also be expanded in powers of $1/\hat \phi$ and the
flow of the ultralocal function $U_{k}^{UL}(\phi_{1})$ is \textit{not} affected by the non-ultralocal contributions. This property generalizes 
to the higher cumulants and to the case where the fields are not uniform in the Euclidean space. This is different from what is encountered 
in the equilibrium case when, studying the asymptotic dominance of the ground state.\cite{tissier11,tissier12b} Along the same lines, the 
flow for any $X_{kn}(\phi)$ is independent of the higher order terms of the expansion. For instance, the flow of $X_{k0}(\phi)$ reads
\begin{equation}
\begin{aligned}
\label{eq_betax0}
&\partial_t X_{k0}(\phi)\vert_{\phi}\equiv \beta_{X_0,k}(\phi) =
 \frac{1}{2(p-1)}\tilde {\partial}_t \int_q \Big \{X''_{k0}(\phi)\times \\&
\big [ \Gamma_{k2}^{(11)}(q^2;\phi,\phi)-\tilde{R}_k(q^2)\big ]  \widehat P_k(q^2)^2+
2X'_{k0}(\phi) \widehat P_k(q^2)^2 \times \\& 
\Big (2\widehat P_k(q^2)\Gamma_{k1}^{(3)UL}(q^2;\phi) 
\big [\Gamma_{k2}^{(11)}(q^2;\phi,\phi)-\tilde{R}_k(q^2)\big ] - \\&
\big [ \Gamma_{k2}^{(21)}(q^2;\phi,\phi)+ \Gamma_{k2}^{(12)}(q^2;\phi,\phi)\big ] \Big )+ 
3 X_{k0}(\phi) \widehat P_k(q^2)^2  \times \\&
\Big(2\widehat P_k(q^2)^2  \Gamma_{k1}^{(3UL)}(q^2;\phi)^2
\big [ \Gamma_{k2}^{(11)}(q^2;\phi,\phi)-\tilde{R}_k(q^2)\big ] -\widehat P_k(q^2) \\&
\times \big [ \Gamma_{k2}^{(21)}(q^2;\phi,\phi)+ \Gamma_{k2}^{(12)}(q^2;\phi,\phi)\big ]
+ 2  \Gamma_{k2}^{(22)}(q^2;\phi,\phi) \Big)\Big \}\,.
\end{aligned}
\end{equation}
If $X_{k0}$ is equal to zero at the microscopic scale $\Lambda$, which is the initial condition for the RG flow ($X_{\Lambda0}=0$),
then it is obvious from the above equation that it stays zero all along the flow. The power $p$ of the leading behavior in $1/\hat\phi$ 
in Eq.~(\ref{eq_expansion_large}) is thus fixed by the initial condition. The latter is a mean-field-like description, which amounts to 
an effective zero-dimensional model. In the following we therefore make a detour to study a toy model: the $d=0$ version of the 
out-of-equilibrium RFIM considered here. This will also prove instructive to elucidate the physics behind the non-ultralocal corrections. 

\subsection{Zero-dimensional RFIM model}
\label{sec:toy}  

We consider the $d=0$ version of the $\phi^4$ theory in a quenched random field defined by Eqs. (1-3) of the main text, {\it i.e.}
\begin{equation}
\label{eq_action_d0}
S(\phi;h+J)=\frac{r}{2}\phi^2+\frac{u}{4!}\phi^4-(J+h)\phi \,,
\end{equation}
where $r<0$, so that the extremization equation $\partial S(\phi;h+J)/\partial\phi=-\vert r\vert \phi+(u/3!)\phi^3-(J+h)=0$ 
has three solutions for a range of $h+J$ around zero. The partition function in the presence of 
an auxiliary field $\hat J$ has contributions from the three solutions (when present):
\begin{equation}
\label{eq_Zd0}
\mathcal Z_h(\hat J,J)=\sum^{3}_{\alpha=1}(-1)^{n_{\alpha}}\mathcal C_{\alpha}(h+J)e^{\hat{J}\phi_{\alpha}(h+J)}
\end{equation}
where $\mathcal C_{\alpha}(h+J)$ is the characteritic function of the interval of $h+J$ over which $\phi_{\alpha}$ exists and $n_{\alpha}$ 
is the index of the $\alpha$th solution (here, $+1$ for a maximum and $-1$ for a minimum).

Consider again for illustration the descending branch of the hysteresis characterized by the extremal state with maximum magnetization 
$\phi_M(h+J)$ which is obtained when $\hat J\rightarrow +\infty$. When $\hat J$ is large but not infinite, the generating functional 
in Eq.~(\ref{eq_Zd0}) is dominated by $\exp[\hat J \phi_{M}(h+J)]$ (the maximal state is a minimum). Corrections that do not vanish 
exponentially with $\hat J$ can only occur for the range of $h+J$ where a second solution has a magnetization, say $\phi_S$, that is 
within $1/\hat \phi$ of $\phi_M$. This takes place in the vicinity of the point $h+J=J^*$ and $\phi=\phi^*$ 
where the extremal state (minimum) collapses 
with the nearby saddle-point (maximum). Then, the disorder average of the logarithm of the generating functional, 
$\mathcal \mathcal W_1(\hat J,J)=\overline{\ln \mathcal Z_h(\hat J,J)}$ is given at leading orders in $\hat J$  by
\begin{equation}
\begin{aligned}
&\mathcal W_1(\hat{J},J)- \hat J\; \overline{\phi_M(h+J)}\sim
\\& \int d(\delta h)\frac{e^{-\frac{(\delta h-J+J^*)^2}{2\Delta_B}}}{\sqrt{2\pi \Delta_B}}
\ln(1-e^{-2\hat{J}\sqrt{\frac{2\vert \delta h\vert }{u\phi^*}}})
\end{aligned}
\end{equation}
where the integral over $\delta h=h+J-J^*$ is restricted to a finite range around $0$. When $\hat J \rightarrow \infty$, this leads to
\begin{equation}
\begin{aligned}
\label{eq_W1d0}
\mathcal W_1(\hat{J},J)- \hat J\; \overline{\phi_M(h+J)}\sim \frac{e^{-\frac{(J-J^*)^2}{2\Delta_B}}}{\sqrt{2\pi \Delta_B}}\frac{1}{\hat J^2}
\end{aligned}
\end{equation}
From the above behavior one immediately obtains that $\hat \phi= \mathcal W_1^{(01)}(\hat{J},J) \sim \hat J + \mathcal O(1/\hat J^2)$, 
$\phi= \mathcal W_1^{(10)}(\hat{J},J)\simeq \overline{\phi_M(h+J)} + \mathcal O(1/\hat J^3)$
and that $\mathsf \Gamma_1(\phi,\hat \phi)=-\mathcal W_1(\hat{J},J)+\hat{J}\phi +J\hat\phi$ is given by
\begin{equation}
\begin{aligned}
\label{eq_Gamma1d0}
\mathsf \Gamma_1(\phi,\hat \phi)= \hat \phi \, J_M(\phi)+ \frac{Y(\phi)}{\hat \phi^2} + \mathcal O(1/\hat \phi^3)
\end{aligned}
\end{equation}
when $\hat \phi,\hat J \rightarrow \infty$, where $J_M(\phi)$ is the inverse function of $\overline{\phi_M(h+J)}$. 

The first term of the right-hand side of Eq.~(\ref{eq_Gamma1d0}) is the contribution that is ultralocal in the Grassmann 
coordinates and the second one is the dominant non-ultralocal correction (the fermionic fields have been set to zero for simplicity).
The latter therefore behaves as $1/\hat\phi^2$ when $\hat \phi\rightarrow \infty$. The same result is valid for the mean-field 
approximation in general dimension $d$ as it essentially amounts to considering a self-consistent zero-dimensional effective 
system. This shows that the non-ultralocal contribution at the UV scale $\gamma_{k=\Lambda1}( \phi, \hat \phi)$ (see 
Eqs.~(\ref{eq_cumulant_1_nonultralocal_components}, \ref{eq_expansion_large}) above) behaves as $1/\hat\phi^3$ 
at large $\hat\phi$, \textit{i.e.} $p=3$.

\subsection{Results: irrelevance of non-ultralocal corrections at large distance}
    
To investigate the long-distance physics in the vicinity of the out-of-equilibrium critical point,  we must cast the functional RG flow 
equations in a dimensionless form by using scaling dimensions appropriate for a zero-temperature fixed point. This is described 
in the main text (and in more detail in Ref.~[\onlinecite{tissier12b}]). Accordingly, we define a dimensionless non-ultralocal 
contribution from $X_{k0}(\phi)=k^{\kappa}\chi_{k0}(\tilde\varphi)$;  the associated beta function in 
Eq.~(\ref{eq_betax0}) similarly scales as 
$\beta_{X_0,k}(\phi)=k^{\kappa}\beta_{\chi_{0},k}(\tilde\varphi)$, so that in dimensionless form,
\begin{equation}
\label{eq_scaled_chi0}
\partial_t \chi_{k0}(\tilde\varphi)\vert_{\tilde\varphi}=-\kappa \chi_{k0}(\tilde\varphi)+\frac 12 (d-4+\bar\eta) 
\tilde\varphi \chi_{k0}'(\tilde\varphi)
 + \beta_{\chi_{0},k}(\tilde\varphi)\,.
\end{equation}

The naive expectation for the scaling of 
$\hat \phi$ is that it behaves as $k^{(d+2\eta-\bar\eta)/2}$. However, $\hat\phi$ should rather be adjusted so that $\hat J$ 
can go to infinity even at the fixed point since this is the way to select the extremal state. As
\begin{equation}
\hat J =\frac{\partial\mathsf \Gamma_{k1}[\phi,\hat \phi]}{\partial \hat \phi} \simeq U_k^{(2)}(\phi)\,\hat \phi
\end{equation}
with $U_k^{(2)}(\phi) \sim k^{2-\eta}$ near the fixed point, $\hat \phi$ should scale as $k^{-(2-\eta)}$. More precisely, we define  
a constant $\hat\phi_0$ which asymptotically behaves as $\hat J$ and such that $\hat\phi$ evolves under the RG flow close  
to the fixed point as $k^{-(2-\eta)}\hat\phi_0$. 
The relevant non-ultralocal quantity to be compared with the ultralocal one, 
$U_k^{(1)}(\phi)=k^{(d-2\eta+\bar\eta)/2}u_k'(\tilde\varphi)$, can thus be expressed as
\begin{equation}
\label{eq_omega}
\frac{X_{k0}(\phi)}{\hat \phi^3}=k^{\omega}\, \frac{\chi_{k0}(\tilde\varphi)}{\hat \phi_0^3}
\end{equation}
with $\omega=\kappa+3(2-\eta)$.

We solve Eq.~(\ref{eq_scaled_chi0}) as an eigenvalue equation by setting $\partial_t \chi_{*0}(\tilde \varphi)=0$ and 
by   using the ultralocal functions already found for the fixed point. 
The 1-replica functions $u_*''(\tilde \varphi)$, $z_*(\tilde \varphi)$, and $\chi_{*0}(\tilde \varphi)$ are discretized on a grid 
of $2\times110+1$ points with a mesh of $\delta\tilde \varphi=0.005$, thus giving the range of the field $\tilde\varphi$ from 
$-0.55$ to $0.55$. The 2-replica function, \textit{i.e.} the second cumulant of the renormalized random field, $\delta_*(x,y)$ 
with $x=(\tilde\varphi_1+\tilde\varphi_2)/2$ and $x=(\tilde\varphi_1-\tilde\varphi_2)/2$, is discretized 
on a trapezoidal grid with a base identical to the domain of the 1-replica functions and a height of $36$ points. The 
mesh in the second field $y$ is identical to that of the field $x$, $\delta y=\delta x$.  (We checked that by doubling 
the resolution of the mesh, our results change on the $4$th digit and by changing the range of fields, the change is on the $6$th digit.)

\begin{figure}[h!]
   \begin{center}
   \includegraphics[width=\linewidth]{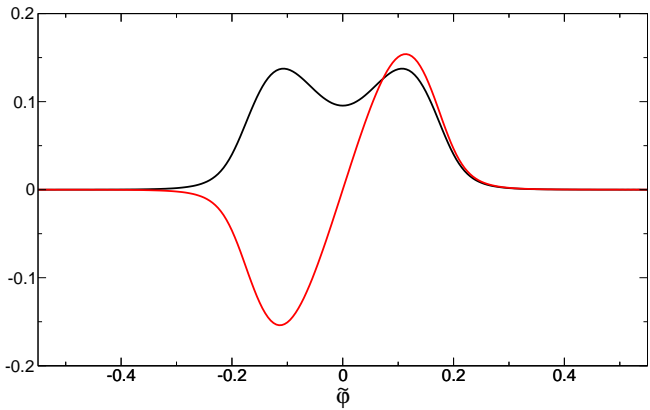}  
    \caption{The two nontrivial solutions $\chi_{*0}(\tilde \varphi)=0$ in $d=4$. One is 
   $Z_2$ symmetric and the other one is antisymmetric. In this particular spatial dimension, the corresponding eigenvalues 
   $\kappa_1$ and $\kappa_2$ are almost identical. }
    \label{fig3}
  \end{center}  
\end{figure}

\begin{figure}[h!]
  \begin{center}
   \includegraphics[width=\linewidth]{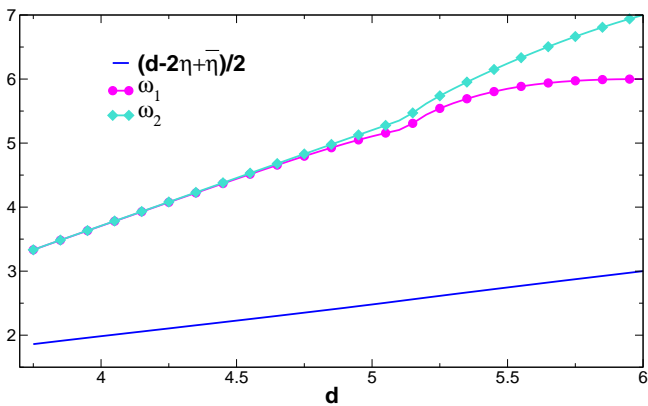}
    \caption{Exponents $\omega_{1}$ (for the $Z_2$ symmetric eigenfunction) and $\omega_{2}$ (for the $Z_2$ 
    antisymmetric eigenfunction) of the nonultralocal correction as a function of dimension $d$. We also plot 
     the scaling exponent $(d-2\eta+\overline{\eta})/2$ of the corresponding ultralocal term (blue line). It is clear that the 
     non-ultralocal terms are subdominant and do not affect then the universal behavior. }\label{fig4}
  \end{center}
\end{figure}

There are two nontrivial solutions of the eigenvalue equation, which are illustrated in Fig.~\ref{fig3} for the dimension $d=4$. 
The eigenvalues $\kappa_{1,2}$ are monotonically increasing functions of the dimension and they reach at the upper critical 
dimension $d=6$ values that can be analytically derived: $\kappa_1(d= 6)= 0$ and $\kappa_2(d= 6)= 1$. The eigenvalue 
$\omega$ defined in Eq.~(\ref{eq_omega}) is then simply obtained by adding $3(2-\eta)$. The result 
is plotted as a function of dimension in Fig. 2. 
One can clearly see that the exponent of the nonultralocal contribution (whether obtained from the symmetric or the antisymmetric 
solution) is much larger than that of the ultralocal term. This proves that the corrections to Grassmann ultralocality are irrelevant 
at large distance, \textit{i.e.} that the selection of the extremal states is properly ensured.

\subsection{Physical interpretation of the non-ultralocal terms}

We conclude this material by briefly discussing the physics behind the non-ultralocal corrections. A first hint is given by the 
zero-dimensional model. As seen in section \ref{sec:toy}, the most significant contribution associated with the corrections come 
from rare situations where the extremal state almost coincides (within $1/\hat J$ when $\hat J\rightarrow +\infty$) with a nearby 
saddle-point.

The reasoning can be carried over to the general case. The non-ultralocal corrections are due to rare events  
where there is an almost degeneracy (within $1/\vert \hat J\vert$) between the relevant extremal state and a nearby 
stationary state, solution of the stochastic field equation $\delta S[\varphi;h+J]/\delta\varphi(x)=0$, 
with a very different configuration yet a very 
close total magnetization. These rare instances make the non-ultralocal contributions vanish at large distance as a power law 
rather than the naively anticipated exponential decay. This is somewhat reminiscent of the role of 
power-law rare ``droplet" excitations near the ground state at low but nonzero temperature in the 
equilibrium case.\cite{villain84,fisher86,tissier08}

\end{document}